\documentclass[aps,prl,reprint,floatfix,longbibliography,twocolumn]{revtex4-2}
\usepackage{amsmath,amssymb,amsthm,hyperref,graphicx}
\hypersetup{hidelinks}
\usepackage[capitalise]{cleveref}
\hypersetup{colorlinks=true, linkcolor=blue, citecolor=red, urlcolor=blue}
\begin{document}
\title{Plasmon-Driven Giant Amplification of Ultrashort Spin Current}
\author{H. Y. Yuan$^{1}$}
\email{Contact author: hyyuan@zju.edu.cn}
\author{Rembert A. Duine$^{2,3}$}
\affiliation{$^{1}$Institute for Advanced Study in Physics, Zhejiang University, 310027 Hangzhou, China}
\affiliation{$^{2}$Institute for Theoretical Physics, Utrecht University, 3584 CC Utrecht, The Netherlands}
\affiliation{$^{3}$Department of Applied Physics, Eindhoven University of Technology, 5600 MB Eindhoven, The Netherlands}

\date{\today}

\begin{abstract}
A key challenge in spintronics is to efficiently generate and manipulate spin current for information processing. 
Here we study ultrashort spin transport and associated terahertz (THz) emission in a hybrid structure comprising gold nanoparticles, a ferromagnet (FM) 
and a normal metal (NM) and show that plasmon excitation in the nanoparticles strongly enhances the electron-magnon scattering 
rate through heating effects, thereby amplifying the spin current generation at the FM$|$NM interface. This effect is even more pronounced when the FM is an insulator 
with a thickness much smaller than the nanoparticle size. In this case, the gold nanoparticle and NM substrate form a nanocavity with the FM as a dielectric layer, trapping plasmons inside the gap. 
The resulting spin current can be amplified by two orders of magnitude as compared to the case without plasmon excitations. Our findings provide 
a novel route to design efficient spintronic THz devices and further open the door to the interdisciplinary field of spintronics and nanophotonics.
\end{abstract}

\maketitle
\textit{Introduction.--}
Spintronics manipulates spin and spin excitations in solid-state systems for information processing. Spin excitations can generate spin currents to carry and transport information \cite{Maekawabook,ChumakNP2015,YuanQM}. One can generate spin current in a magnetic system by microwave antennas, electric injection, mechanical strain, temperature gradients, ultrafast lasers etc. Of particular interest is that ultrashort spin currents generated by a femtosecond laser pulse can be converted into an ultrashort electric pulse, emitting terahertz (THz) 
electromagnetic waves in a ferromagnet$|$normal-metal (FM$|$NM) heterostructure \cite{KampNN2013, SeifertNP2016, YangAOP2016, WuAM2017, Evangelos2020, SeifertAPL2022}. Compared to the traditional THz emitters based on semiconductors \cite{BerryNC2013,DreyAPL2005}, spintronic emitters offer advantages such as ultrabroad 
bandwidth, high efficiency and geometric flexibility. Despite some promising results in the last few years, this field remains a major open 
research field on both the fundamental and application sides. One key issue is how to efficiently generate strong THz emission \cite{Evangelos2020,SeifertAPL2022}. Current strategies focus on optimizing the laser absorption rate of the magnetic layer, increasing the spin-to-charge conversion efficiency by using metals with high spin-Hall angles, and combining the forward and backward pulses \cite{SeifertAPL2022}. However, all these improvements operate within existing frameworks and may face inherent limitations from material and interface constrains. It thus needs to explore novel mechanisms and structures for generating ultrashort spin currents with strong THz emission.

In general, two mechanisms compete to explain the ultrashort spin current generation in the FM$|$NM hybrid. First, the pump laser generates super-diffusive spin transport in the magnet and launches spin current toward the metal layer \cite{BattiatoPRL2010, BattiatoPRB2012}. Second, heating due to pump laser induces angular momentum transfer between magnons and electrons through spin-flip scattering process and generates a non-equilibrium spin accumulation at the interface, pumping a spin current into the normal metal \cite{ChoiNC2014,TvetenPRB2015,BeensJPCM2023,BarbeauPRB2023}. Both mechanisms depend on the electromagnetic energy or electric field of the pump pulse. Intuitively, one can enhance the spin currents by amplifying the electric fields at the magnetic surface. On the other hand, it is widely known that metallic nanoparticles (NPs) can strongly trap the light at sub-wavelength scale and thus enhance the near-field distribution via localized surface plasmons (LSPs) \cite{MaierBook,Willets2007,MayerReview2011}.  Uchida \textit{et al.} observed spin pumping driven by plasmons at the FM$|$NM interface \cite{UchidaNC2014}. Zhang \textit{et al.} found the enhancement of THz emission in a Co$|$Pt heterostructure with gold nanoparticles deposited at the magnetic surface, but the mechanism remains unclear \cite{ZhangAPL2024}. Conversely, Kuznetsov \textit{et al.} reported the suppression of the spin wave signal by surface plasmons \cite{KuznetsovSA2025}. 


In this Letter, we study laser-induced giant enhancement of spin transport in hybrid gold nanoparticles (GNPs), ferromagnet and normal metal structure. For metallic FMs, we show that GNPs strongly modify the electric field distribution at the FM surface, boosting the electron-magnon scattering rate and ultrashort spin current. The spin current can be two times larger than that without GNPs, depending on the incident laser power and particle size. This explains the recent experimental results \cite{ZhangAPL2024}. For ultrathin insulating FMs like yttrium iron garnet (YIG), GNPs and the NM substrate form a nanocavity, which can trap light. Such strong electric fields significantly increase the electron temperature of the metal layer and then enhance the scattering rate of electrons and spins at the interface, amplifying the spin current up to two orders of magnitude. Our results should open a novel route to design efficient spintronic THz emitters while the plasmonic nanoparticles may provide a novel knob to control the spin excitations and stimulate the interdisciplinary studies of spintronics and plasmonics.

\begin{figure}
	\centering
	\includegraphics[width=0.48\textwidth]{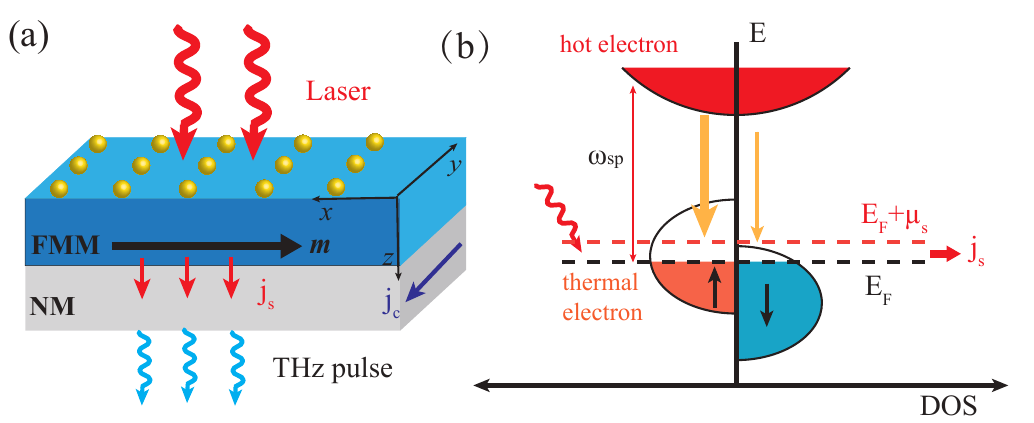}\\
	\caption{(a) Scheme of GNPs/FMM/NM hybrid structure. An incident laser excites the surface plasmon resonance in the GNPs and enhances the spin current generation at the FMM$|$NM interface. (b) Physical picture of the plasmon-induced spin current generation. DOS is short for density of states.}\label{fig1}
\end{figure}

\textit{GNPs$|$FMM$|$NM heterostructures.}---We consider a GNPs, ferromagnetic metal (FMM) and NM hybrid structure as shown in Fig. \ref{fig1} (a). The physical picture of plasmon-induced spin current enhancement is illustrated in Fig. \ref{fig1}(b). An incident laser excites the LSPs on the GNPs, which then generates strong electric field at the FMM surface, exciting the thermal electrons to form a hot electron gas. Due to the electron-electron and electron-environment interactions, the hot electrons relax back to the thermal state, enhance the electron-magnon scattering rate and then create a non-equilibrium spin chemical potential $\mu_s$, driving the spin currents. Compared to systems without GNPs, the electron-magnon scattering rate and hence the spin current are enhanced due to the plasmon-induced electric field amplification at the magnetic surface.

Let us follow this picture to quantify the enhancement of spin current. The electron transport in the ferromagnetic layer is modeled by the Boltzmann equation
\begin{equation}\label{Boltzman equaton time domain}
\frac{\partial n_\eta(\mathbf{r},v,t)}{\partial t} + v\frac{\partial n_\eta(\mathbf{r},v,t)}{\partial z} =A_\eta(\mathbf{r},v,t)-\frac{n_\eta(\mathbf{r},v,t)}{\tau_\eta},
\end{equation}
where $n_\eta(\mathbf{r},v,t)$ and $\tau_\eta$ are respectively the electron density and relaxation rate in the magnet with polarization $\eta=\uparrow,\downarrow$, $v$ is electron speed along the thickness direction. The source term $A_\eta(\mathbf{r},v,t)$ originates from the laser. For simplicity, we assume that the source term is $\delta$-function like at the magnetic surface, i.e. $A_\eta(\mathbf{r},v,t)=a_\eta A_{0}(t)E^2(x,y)\delta(z)$, where $a_\eta$ is the energy absorption efficiency of the magnet, $A_{0}(t)=\exp(-t^2/\sigma^2)$ describes the temporal profile of laser with $\sigma$ being the laser pulse width. $E(x,y)$ represents the spatial distribution of the electric field. In the absence of metallic particles, $E(x,y)$ is independent on $x$ and $y$ and is equal to the electric field of the incident light, i.e. $E(x,y)=E_0$. The metallic particles strongly modify the electric field at the surface of the magnetic layer, as we discuss now.

\begin{figure}
	\centering
	\includegraphics[width=0.48\textwidth]{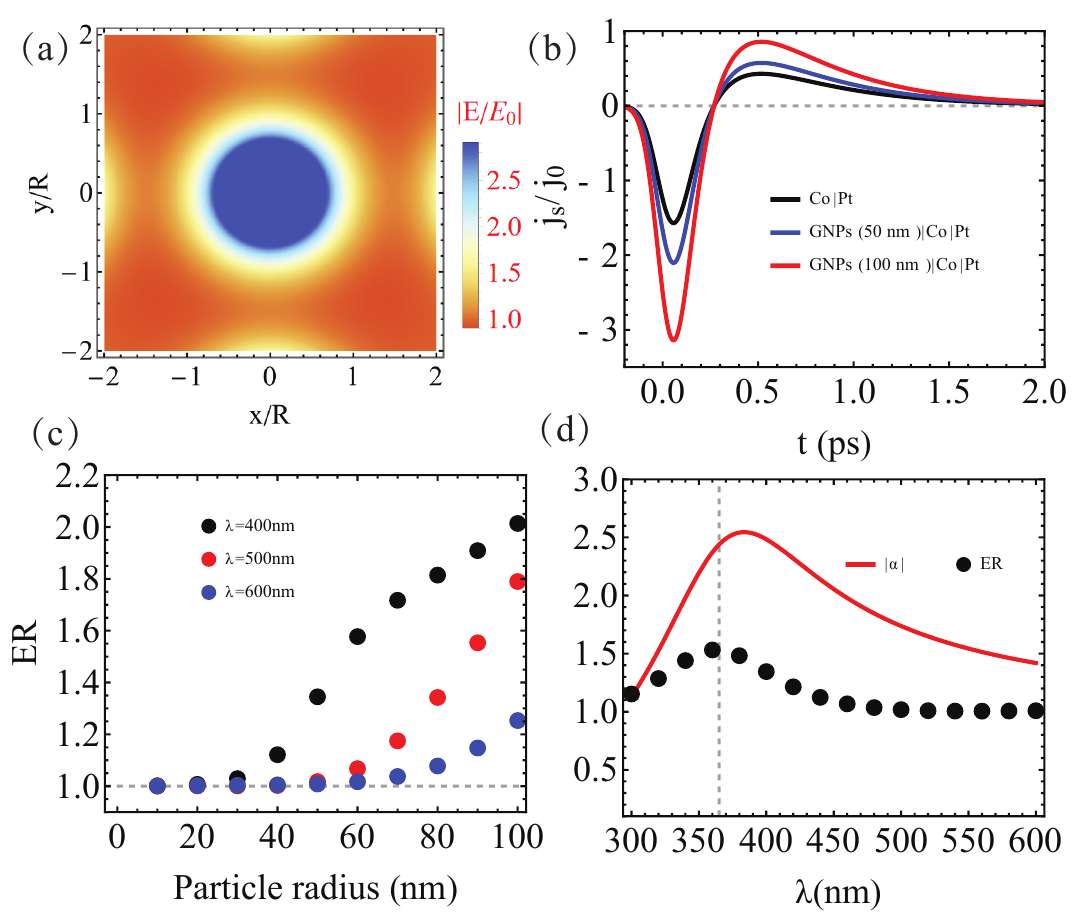}\\
	\caption{(a) Electric field distribution across the interface. Parameters of gold are used. $\omega_\mathrm{sp}=1.05 \times 10^{16}~\mathrm{Hz},~\omega_\tau=4.07\times 10^{13}~\mathrm{Hz},~\epsilon_\infty=1.53$ \cite{Ordal1983,Pereira2013}. Particle density $n_\mathrm{Au}=2.78~\mathrm{\mu m}^{-2}$. (b) Time evolution of spatially averaged spin current with and without gold nanoparticles. $\sigma=0.1~\mathrm{ps},~\tau_e=0.45~\mathrm{ps},~\tau_m=0.15~\mathrm{ps}$ \cite{KoopmansPRL2005}. (c) Particle size dependence of the enhancement factor for laser wavelength $\lambda=400$, $500$ and $600$ nm, respectively. (d) Laser wavelength dependence of the enhancement rate and gold polarizability, the particle radius $R=50$ nm.}\label{fig2}
\end{figure}

For a metallic sphere, the incident light will excite the localized surface plasmons inside the sphere, characterized by an electric dipole $\mathbf{p}$. The general theory was derived by Mie which has a complex dependence of the geometry of the particles \cite{Bohren1983}. For small spherical particles, the leading order of the electric dipole is approximated as $\mathbf{p}= \epsilon_0 \epsilon_m \alpha \mathbf{E}_0 e^{-i\omega t}$, where $\epsilon_0$ and $\epsilon_m$ are respectively the permittivity of vacuum and the surrounding medium, and the polarizability $\alpha$ reads \cite{Meier1983,KuwataAPL2003}
\begin{equation}\label{polarizability}
\alpha = \frac{4\pi R^3(1-\frac{1}{10}(\epsilon+\epsilon_m)x^2)}{\frac{\epsilon+2\epsilon_m}{\epsilon-\epsilon_m} - (\frac{\epsilon}{10}+\epsilon_m)x^2 - \frac{10}{3}i \epsilon_m^{3/2}x^3},
\end{equation}
where $R$ is the radius of the sphere, $x=kR$ with $k$ being the wavevector of incident light. $\epsilon =\epsilon_\infty -\omega_\mathrm{sp}^2/(\omega^2+i\omega_\tau \omega)$ is the electric permittivity of the metal, with $\epsilon_\infty$, $\omega_\mathrm{sp}$, $\omega_\tau$ being the permittivity at infinite frequency, plasmon frequency and electron relaxation rate of the metals, respectively. Here the first and second terms in the denominator of Eq. \eqref{polarizability} represent the contributions of electric dipole and quadruple, respectively. The third imaginary term $x^3$ characterizes the absorption of light by the nanoparticles and will become important for larger particles. In the Rayleigh limit, $kR \ll 1$, we recover the well known result of metallic polarizability $\alpha = (\epsilon-\epsilon_m)/(\epsilon+2\epsilon_m)$. We assume a square lattice of gold spheres distributed on the magnetic surface with density $n_\mathrm{Au}$, as shown in Fig. \ref{fig1}(a). The total electric field on the magnetic surface is a superposition of all the electric dipoles's contribution \cite{Jacsonbook}, i.e.
\begin{equation}\label{surface-Efield}
\begin{aligned}
\mathbf{E}(x,y)=\mathbf{E}_0 + \frac{1}{4\pi \epsilon_0} \sum_i k^2(\mathbf{n} \times \mathbf{p}_i) \times \mathbf{n}\frac{e^{ikr}}{r} \\
+ \frac{1}{4\pi \epsilon_0} \sum_i [ 3 \mathbf{n} (\mathbf{n}\cdot \mathbf{p}_i) - \mathbf{p}_i ] \left ( \frac{1}{r^3} - \frac{ik}{r^2} \right )e^{ikr},
\end{aligned}
\end{equation}
where the first, second and third terms refer to the incident field, near-zone and far-zone electric field respectively. The directional vector $\mathbf{n}=\mathbf{r}/|\mathbf{r}|$ with $\mathbf{r}=(x,y,z)$. Figure \ref{fig2}(a) shows the typical electric field distribution near the bottom of one particular sphere. Clearly, the strength of the electric field is enhanced and such an enhancement decays as one goes away from the bottom center.
We will show that the modified electric field distribution will enhance the average electron-magnon scattering rate in the magnetic layer and then amplify the generation of ultrashort spin current.

To proceed, we solve the dynamic equation \eqref{Boltzman equaton time domain} in frequency space and derive the electron density as
\begin{equation}
n_\eta(\mathbf{r},v_\eta,\omega) = \frac{a_\eta A_0(\omega)}{v}E^2(x,y)\exp \left [-\frac{z}{v\tau_\eta} (1+i\omega \tau_\eta)\right ].
\end{equation}
Note that the electron velocity is randomly distributed over a halfsphere with speed $v_\eta$, i.e. $v=v_\eta \cos \theta$. Here we first average over the solid angle of the halfsphere and derive a velocity-direction-independent electron density. Further, we  integrate over the whole sample and derive the average electron density as
\begin{equation}
n_\eta(\omega) = \frac{\pi a_\eta A_0(\omega)}{2v_\eta}\langle E^2 \rangle\left [ K_0(z_\eta)L_{-1}(z_\eta)+K_1(z_\eta)L_0 (z_\eta)\right ]
\end{equation}
where $z_\eta=(1+i\omega \tau_\eta)d/(v_\eta \tau_\eta)$. $K_n(z_\eta)$ and $L_n(z_\eta)$ are respectively the modified second-kind Bessel function and Struve function. $\langle E^2 \rangle$ is the spatial average of electric field at the magnetic surface.

Based on spin angular momentum conservation, the non-equilibrium electron spin density $\delta n_e$ close to the Fermi surface satisfies the equation
\begin{equation}\label{njs_continue_eq}
\frac{\partial \delta n_e(t)}{\partial t} + \frac{j_s(t)}{d} = -2 I_{sd}(t) - \frac{\delta n_e (t)}{\tau_s},
\end{equation}
where $j_s$ is the magnitude of the spin current propagating in the $z$-direction, $I_{sd}$ is the electron-magnon scattering rate and characterizes the angular momentum transfer between magnons and electrons while $\tau_s$ is the spin relaxation time of thermal electrons. On the other hand, both the electron spin density $\delta n_e$ and electron spin current are connected to the spin accumulation $\mu_s$. The spin density $\delta n_e=D_s \mu_s$ with $D_s$ being the spin density of states at the Fermi surface while $j_s = g_s/\hbar \mu_s$ \cite{CornePRB2016} with $g_s$ being the interfacial spin conductance.
We solve Eq. \eqref{njs_continue_eq} in frequency space and obtain
\begin{equation}
j_s(\omega)= \frac{-2dI_{sd}(\omega)}{1+(1+i\omega \tau_s) \tau_g/\tau_s},
\end{equation}
where $\tau_g \equiv g_s/(\hbar D_s d)$.

Here the electron-magnon scattering rate $I_{sd}$ characterizes the time dependence of the magnon temperature as $C_{n,T} \partial_t T_m= I_{sd}(t)$ \cite{BeensJPCM2023} with $C_{n,T}$ being the magnonic specific heat. Within a linear approximation, $I_{sd}$ can be expressed as the temperature difference between electron and magnon subsystems
$I_{sd} = \frac{C_{n,T}}{\tau_m} (\delta T_e - \delta T_m) $ \cite{Xiao2010,BeensJPCM2023} with $\tau_m$ being the typical demagnetization rate. Then we can eliminate $T_m$ and solve for the electron temperature as
$I_{sd}(\omega) = i\omega C_{n,T}/(1+i\omega \tau_m)\delta T_e(\omega)$. The change of the electron temperature thus satisfies the equation
\begin{equation}\label{electron temperature equation}
C_e\frac{\partial \delta T_e}{\partial t} = \hbar \omega_{ph} \sum_{\eta} \frac{n_\eta}{\tau_\eta} - C_e\frac{\delta T_e}{\tau_e},
\end{equation}
where $C_e$ is the electronic specific heat. By solving this equation, we derive the change of electron temperature as $\delta T_e (\omega)= \tau_e/(1+i\omega \tau_e) \hbar \omega_{ph}/C_e\sum_{\eta} n_\eta (\omega)/\tau_\eta$. Because the spin of hot electrons relaxes very fast, i.e. $\omega \tau_\eta \ll 1$, we may approximate $\sum_{\eta} n_\eta(\omega)/\tau_\eta $ by its zero frequency component, i.e. $\sum_{\eta}  n_\eta(\omega)/\tau_\eta = A_0(\omega) \langle E^2\rangle \Theta_0 $ with $\Theta_0 =\pi a_\eta/(2v_\eta \tau_\eta) K_0(z_\eta^0)L_{-1}(z_\eta^0)+K_1(z_\eta^0)L_0 (z_\eta^0)$ with $z_\eta^0=d/v_\eta \tau_\eta$. By substituting $\delta T_e (\omega)$ into $I_{sd}(\omega)$ and taking account of Eq. (5), we express the electron-magnon scattering rate as
\begin{equation}
I_{sd}(\omega) = \frac{\hbar \omega_{ph}\Theta_0 C_{n,T} \langle E^2 \rangle (i\omega \tau_e) A_0(\omega)}{2C_e(1+i\omega \tau_e)(1+i\omega \tau_m)}.
\end{equation}

After a Fourier transform, we find for the time evolution of the spin current
\begin{equation}\label{jstmetal}
j_s(t) = j_0 \frac{\langle E^2 \rangle}{E_0^2}   \left [ \varrho(t,\tau_e) - \frac{\tau_e}{\tau_m} \varrho (t,\tau_m) \right ],
\end{equation}
where $j_0 =\sqrt{2} \sigma C_{n,T}\hbar \omega_{ph}\Theta_0 E_0^2/(1+\tau_g/\tau_s)(\tau_e-\tau_m)$ is the amplitude of the spin current without GNPs and
\begin{equation}
\varrho(t,\tau_e) = \exp \left(\frac{\sigma^2-4t\tau_e}{4\tau_e^2} \right) \left [ 1+\mathrm{Erf} \left(\frac{t}{\sigma} -\frac{\sigma}{2\tau_e}\right ) \right ]
\end{equation}
with $\mathrm{Erf}$ being the error function.

We proceed by numerically calculating the spatially averaged electric field and the associated spin current according to Eqs. \eqref{surface-Efield} and \eqref{jstmetal}. Figure \ref{fig2}(b) shows the time evolution of the spatially averaged ultrashort spin current with and without nanoparticles for the Co$|$Pt hybrid structures. Clearly, the amplitude of the spin current is enhanced by a factor of two when GNPs are deposited on the magnetic layer and the enhancement is even stronger for larger particle size. To quantify the enhancement, we define the enhancement rate (ER) as the ratio of maximum spin current with and without the GNPs. Figure \ref{fig2}(c) shows that the enhancement rate increases with the particle size up to $R=100$ nm. This is because a larger particle has a larger polarization $|\alpha|$ and thus enhances the electric field distribution at the FM surface. The enhancement rate agrees very well with the experiment using GNPs with an average grain size of around 100 nm \cite{ZhangAPL2024}. Furthermore, as the particle size increases and becomes comparable to the wavelength of the light, the energy absorbtion of the particles also becomes important, characterized by the imaginary term in Eq. \eqref{polarizability}. Then the surface electric field and the spin current enhancement will be suppressed, which explains the saturation behavior for large particle size.  Figure \ref{fig2}(d) shows laser wavelength dependence of the enhancement rate and particle polarizability for $R=80$ nm. The enhancement is correlated to the plasmon excitation, characterized by the magnitude of polarizability  $|\alpha|$. However, the position of maximum polarizability is not always coincident with the maximum spin current enhancement. This is because the surface electric field has a non-monotonic dependence on the laser wavevector $k$ in Eq. \eqref{surface-Efield} besides the influence of particle polarization characterized by the polarizability. The maximum enhancement of the electric field depends on the competition between these two mechanisms.

\begin{figure}
	\centering
	\includegraphics[width=0.48\textwidth]{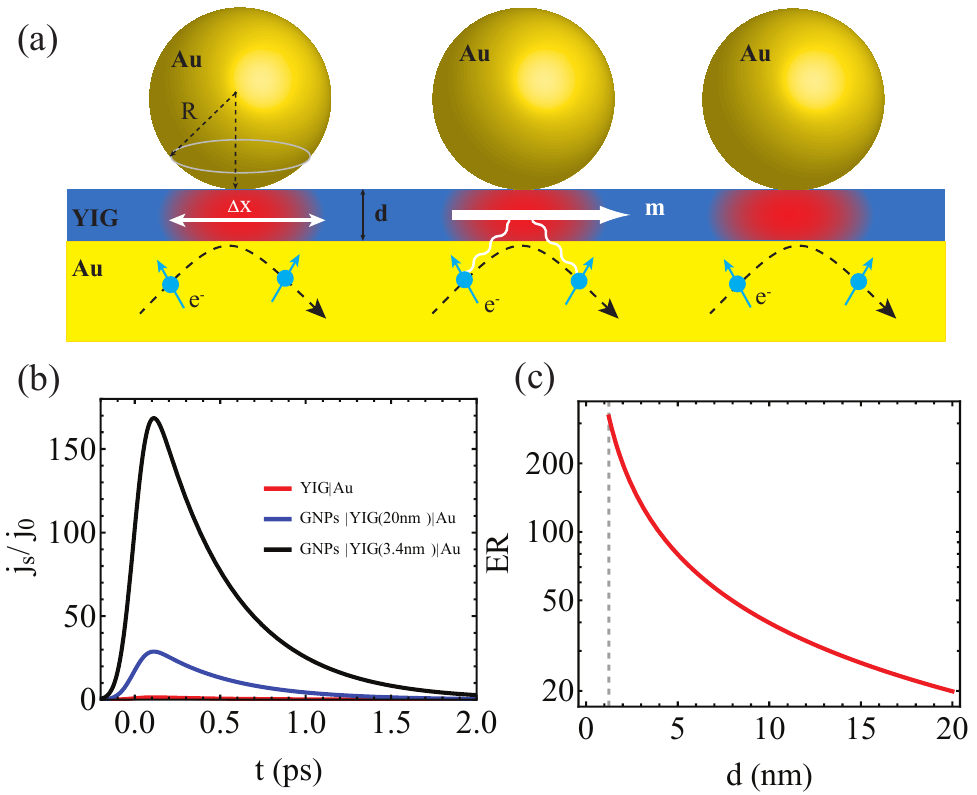}\\
	\caption{(a) Scheme of GNPs/YIG/Au structure. Here the YIG layer is so thin that a nanocavity formed between the GNPs and Au film, which greatly enhances the electric fields inside the cavity. (b) Time evolution of the spin current without and with gold nanoparticles. Parameters are $R=100~\mathrm{nm}$,  $|\alpha|/(4\pi R^3) = 2.5$, $\sigma=0.1~\mathrm{ps},~\tau_e=0.45~\mathrm{ps}$ \cite{KoopmansPRL2005}. (c) Enhancement of the spin current as a function of the thickness of magnetic insulator. The dashed line at $d=1.24$ nm indicates the lattice constant of YIG.}\label{fig3}
\end{figure}

\textit{GNPs$|$FMI$|$NM heterostructures.}---Now, let us consider a hybrid structure of gold nanoparticles, ferromagnetic insulator and nonmagnetic metal, as shown in Fig. \ref{fig3}(a). For ultrathin ferromagnetic insulators (FMIs) like YIG, the gold nanoparticle and the gold substrate form a nanocavity, which can strongly localize the electromagnetic field in between \cite{Romero2006,Aubry2011,Ciraci2012,Savage2012,Baumberg2019}. It has been shown that the localization length of the electric field is $\Delta x = \sqrt{2Rd/\epsilon_d}$ with $\epsilon_d$ being the permittivity of the dielectric and the concentration volume of electric field underneath the sphere is $V_c=  \pi d \Delta x^2/(4\mathrm{ln2})$ \cite{Savage2012,Baumberg2019}. By equalizing the total energy $E_t= \epsilon_0 \epsilon_d E_{m}^2 V_c$ to the total electric dipole energy $\epsilon_0 \alpha E_0^2$, one derives the enhancement of the electric field as \cite{Baumberg2019}
\begin{equation}
\left ( \frac{E_m}{E_0}\right )^2 =8 (\mathrm{ln}2) Q \sqrt{\epsilon_d} |\alpha| \frac{R^2}{d^2}.
\end{equation}
where $Q$ is the effective quality factor of the cavity. For typical parameters of $\mathrm{GNPs|YIG|Au}$ with an ultrathin YIG layer, $R=100~\mathrm{nm},~d= 3.4~\mathrm{nm}$,~$|\alpha|/(4\pi R^3) = 2.5,~\epsilon_d=12.5$ \cite{Lal1998}, we have a huge enhancement of the surface electric field $E_m/E_0 = 8.0 \times 10^2$, which is two orders of magnitude larger than that shown in Fig. \ref{fig1}(a).

The enhanced electric field will also enhance the spin current at the YIG$|$Au interface while the physical mechanism is different from the electron-magnon scattering in a metallic magnet. In the case of a magnetic insulator, a hot electron in the metal first scatters with a magnetic moment and generates a random field on the magnetization. The disturbed magnetic moment then acts backs on the electron within the correlation time $\tau_N$, making it spin-polarized \cite{AdachiRPP2013}. This is a second-order effect and is determined by the spin correlation function of electrons in the NM layer. Within linear response theory, it has been shown that the spin correlation function is connected to the electron temperature and thus the resulting spin current at the interface is $j_s(t) = \int dt' \kappa(t-t')(T_e(t)-T_I)$ \cite{SeifertNC2018}, where $T_e$ is the electron temperature of normal metal depending on the enhanced electric field, $T_I$ is the temperature of the magnetic insulator, and $\kappa(t-t')$ is the response function. Based on this physical picture, the time evolution of the response function is extremely fast, typically on the time scale of $\tau_N\sim a_\mathrm{YIG}/v_F=0.9$ fs for lattice constant  $a_\mathrm{YIG}=1.24$ nm \cite{RuckPRB2014}, and Fermi speed of gold $v_F=1.38 \times 10^6~\mathrm{m/s}$ \cite{GallJAP2016}. Therefore we can approximate the response function as a $\delta$ function and simplify the spin current as $j_s(t)=\kappa(T_e(t) - T_I)$. Here the insulator's temperature does not change significantly on the picosecond time scale due to its low thermal conductivity as compared to metals \cite{OrtizPRM2021,BoonaPRB2014}. Using the calculated electron temperature following Eq. \eqref{electron temperature equation}, we derive the spin current as
\begin{equation}
j_s(t) = j_0 \frac{\langle E^2 \rangle}{E_0^2} \exp\left (-\frac{t}{\tau_e}\right ) \left[1+ \mathrm{Erf}\left(\frac{t}{\tau_e}- \frac{\sigma}{2\tau_e}\right )\right]
\end{equation}
 where $j_0 = \sqrt{2} \pi \sigma \hbar \omega_{ph} \Theta_0 E_0^2 \exp(\sigma^2/4\tau_e^2)$ is the magnitude of the spin current without GNPs. Since the electric field is strongly localized in the region $\Delta x$ below each sphere, we may neglect the electric field outside this region and analytically evaluate its spatial average as $\langle E^2 \rangle =n_\mathrm{Au} E_m^2 V_c/d$.

Figure \ref{fig3}(b) shows the time evolution of spin current with and without GNPs. Clearly, the spin current can be enhanced tremendously when GNPs and gold film form a cavity with YIG as the dielectric layer. The enhancement rate increases with the decrease of the YIG thickness, as shown in Fig. \ref{fig3}(c). This is because a thinner YIG results in a stronger electric field confinement and a larger increase of the electron temperature and spin current at the interface. The maximum enhancement rate using ultrathin YIG of around 3.7 nm thin \cite{Mendil2019,WeiNM2022} is two times larger than that without the GNPs. Such an enhancement makes the efficiency of spin current generation based on magnetic insulators comparable to that of all metallic structures \cite{SeifertAPL2022}.

\textit{Discussions and conclusions.}---We focused on the amplification of spin current induced by surface plasmons in a hybrid spintronic-plasmonic structure. It is straightforward to go a step further to calculate the charge current generated by the spin current through inverse spin Hall effect in the normal metal layer as $\mathbf{j}_c =\theta_\mathrm{SH} (j_se_z) \times \mathbf{m} =-\theta_\mathrm{SH} j_s e_y$ \cite{KampNN2013} with $\theta_\mathrm{SH}$ the spin Hall angle of the normal metal. One quickly finds that the charge current and the associated THz emission will be amplified by the same magnitude as the spin current. We expect that such an enhancement is universal, and does not rely on the specific theoretic model. Within the framework of the super-diffusive model, the emitted THz pulse linearly depends on the pump-pulse fluence, i.e. square of the electric fields \cite{SeifertAPL2022}. The surface plasmon will also enhanced the laser fluence and the subsequent spin current and THz emission.

In summary, we have shown that the localized surface plasmons can enhance the electron-magnon scattering in the hybrid FM$|$NM structures. The enhancement rate depends on the laser intensity, size of the nanoparticle as well as the thickness of magnetic layer, and reaches two order of magnitude larger when the nanoparticle and bottom non-magnetic metal forms a ultrathin nanocavity. Our work explains the existing experimental results, and further advances the design of hybrid plasmonic and spintronic devices as efficient THz emitters.


{\it Acknowledgments.}--- H.Y.Y acknowledges the helpful discussions with Lichuan Jin, Pan Wang, Alireza Qaiumzadeh, and Bert Koopmans. This work is supported by the National Key R$\&$D Program of China (2022YFA1402700).

{}

\end{document}